\documentclass[twocolumn,prl,showpacs]{revtex4}

\usepackage{amsmath}
\usepackage{graphics}
\usepackage{epsfig}

\begin{document}

\title{Limits on Universality in Ultracold Three-Boson Recombination}  
\author{J. P. D'Incao, H. Suno, and B. D. Esry}
\affiliation{Department of Physics, Kansas State University,
Manhattan, Kansas 66506} 

\begin{abstract}
The recombination rate for three identical bosons has been calculated
to test the limits of its universal behavior.  It has been obtained
for several different collision energies and scattering lengths $a$ up
to $10^5$~a.u., giving rates that vary over 15 orders of magnitude.
We find that universal behavior is limited to the threshold region 
characterized by $E \lesssim \hbar^2/2\mu_{12}a^2$, where $E$ is the
total energy and $\mu_{12}$ is the two-body reduced mass.  The
analytically predicted infinite series of resonance peaks and
interference minima is truncated to no more than three of each
for typical experimental parameters. 
\end{abstract}
\maketitle

The development of Bose-Einstein condensates (BECs) with tunable
properties makes possible condensates with a wide range of interaction
strengths. Several experiments
\cite{Inouye,Courteille,Roberts,Weber,Kevrekidis} have investigated
these properties by changing the atomic interactions with an 
external magnetic field near a diatomic Feshbach resonance.  As is
well-known, these interactions are characterized at low temperatures by
the two-body scattering length $a$, which covers the full continuum of
positive and negative values near the Feshbach resonance.  In
fact, all of the essential properties of BECs are determined by the 
scattering length.

Because the two-body scattering length is the only relevant parameter,
the precise shape of the two-body potential does not matter for the 
many-body physics.  One of the remarkable
results to emerge from recent work on ultracold collisions
is that this property holds even for the low temperature {\em
  three}-body physics. This shape independence has allowed theorists
to choose any convenient two-body potential that reproduces the
desired scattering length. We will again exploit this freedom in the
present work to study three-body recombination. Three-body
recombination is the process by which three free atoms collide to form
a diatomic molecule and an unbound atom, setting free enough kinetic
energy to make both atom and molecule escape from typical traps. 

The universal behavior of the three-body recombination rate makes
it possible to derive analytical expressions for
$a>0$~\cite{Nielsen,Esry02,Braaten01} and for $a<0$~\cite{Braaten02}.
In the former case, theory predicts an infinite number of minima in
the rate as the scattering length goes to positive infinity; and in
the latter case, an infinite number of maxima as the scattering length
goes to negative infinity.  The physics behind both features is
closely related to the Efimov effect~\cite{Efimov}. In fact, it has
been suggested that measuring the recombination rate while tuning
through a Feshbach resonance might make possible some of the first
direct experimental evidence of this intriguing effect.

Tuning through such Feshbach resonances can dramatically limit the
density and lifetime of BECs, however, since the three-body
recombination rate was predicted
\cite{Esry01,Nielsen,Esry02,Braaten01,Braaten02} --- and recently
verified experimentally \cite{Weber} --- to increase with the
scattering length as $a^4$.  More recently, three-body recombination
has been used to create composite bosons by pairing fermions in
ultracold gases \cite{Cubizolles,Jochim}. The ultimate goal of this
endeavor has recently been achieved with the observation of
Bose-condensed pairs of fermion atoms~\cite{Jin}.  Despite its importance and recent
advances, much work remains for the theory of three-body
recombination. In particular, the universal behavior
of the recombination rate has not yet been tested by accurate
calculations. 

In this Letter, we show that the recombination rate for identical bosons
is universal only for collision energies in the threshold
regime. Generically, the threshold regime is characterized by
$k|a|\lesssim 1$, or equivalently, when the collision energy is the
smallest energy in the system. For positive scattering lengths, the
energy scale is set by the two-body binding energy; for negative
scattering lengths, by the height of a potential barrier or, in some
cases, by a two-body shape resonance.  Therefore, for a fixed total 
three-body energy $E$, the relation $E\lesssim E_{12}$
indicates when the system is in the threshold regime
where universal behavior is expected.  In this expression,
$E_{12}=\hbar^2/2\mu_{12} a^2$ is the two-body binding energy and
$\mu_{12}$ is the two-body reduced mass.

The experimental consequences of restricting the range
of universal behavior are
striking.  At any nonzero temperature, rather than observing an
{\it infinite} series of resonances or minima, only a finite number of either
will be observable as the scattering length is scanned from $-\infty$
to $+\infty$ --- and even those will be washed out.  For instance, at
1~nK, only three resonances and three minima can be hoped to be seen.

We will further show that the analytical formulas derived in 
Refs.~\cite{Nielsen,Esry02,Braaten01,Braaten02} hold only at zero
energy.  At finite energies, when the scattering length is tuned out
of the threshold regime ($E>E_{12}$), the analytical formulas break
down because they do not take proper account of three important
finite energy effects: unitarity, thermal averaging, and higher
partial waves. Unitarity limits the rate to finite values at finite
temperatures for large scattering lengths, and leads to a saturation 
effect~\cite{Weber,Chris}. Thermal averaging takes account of the
fact that experiments are performed at fixed temperature rather than
fixed collision energy, and higher partial waves must always be
included, in principle. A generalized Wigner threshold
law~\cite{Esry04} guarantees that the $J^\pi=0^+$ contribution
dominates at threshold, where $J$ is the total orbital angular
momentum and $\pi$ is the overall parity.  The next leading
contribution, $2^+$, grows with energy as $E^2$ and with scattering
length as $a^8$, and can quickly become comparable to the $0^+$ rate.

We obtain the recombination rates by solving the Schr{\"o}dinger equation numerically using
the adiabatic hyperspherical representation (see
Refs.~\cite{Esry02,Suno,Esry03,Esry04} for details of our
implementation).  The key to this approach is that the dynamics of the
three-body system are reduced to the motion on a set of coupled
effective potentials that depend only on the hyperradius $R$. The
hyperradius is a collective coordinate that represents, in some sense,
the overall size of the system.  The effective potentials are
determined by solving the adiabatic equation 
\begin{equation*}
H_{\rm ad} \Phi_\nu(R;\Omega) = U_\nu(R) \Phi_\nu(R;\Omega)
\end{equation*}
where $\Omega$ denotes the five hyperangles representing all degrees
of freedom besides $R$.  The adiabatic Hamiltonian $H_{\rm ad}$
includes the kinetic energy for these hyperangles as well as all
interactions.  The effective potentials $U_\nu(R)$ are then used in
the radial equations (atomic units will be used unless otherwise
noted),
\begin{equation}
\left(-\frac{1}{2\mu}\frac{d^2}{dR^2}+U_{\nu}\right)F_{\nu}
-\frac{1}{2\mu}\sum_{\nu'} W_{\nu\nu'} F_{\nu'}=E F_\nu, 
\label{radeq}
\end{equation}
where $F_{\nu}$ is the hyperradial wave function, $E$ is the total
energy, and the three-body reduced mass $\mu$ is related to the atomic 
mass $m$ by $\mu=m/\sqrt{3}$.  The nonadiabatic coupling $W_{\nu\nu'}$
is responsible for inelastic transitions such as three-body
recombination. The effective potentials give a very intuitive picture
for these complicated systems. Moreover, the calculations can be made
as accurate as desired by including more channels in the equation
above (all rates quoted here are accurate to at least three digits and
were obtained with seven channels). The radial equations
(\ref{radeq}) are solved using the variational $R$-matrix method
\cite{Aymar} in order to extract the $S$-matrix.

The three-body recombination rate $K_{3}$ is defined in terms of the
$S$-matrix as~\cite{Esry02,Suno,Esry03,Esry04}
\begin{equation}
K_{3}=\sum_{J,\pi}\sum_{i,f}\frac{192 (2J+1)\pi^{2}}{\mu
k^4}|S^{J\pi}_{f\leftarrow i}|^{2},
\label{K3Def}
\end{equation}  
where $k=\sqrt{2\mu E}$ is the hyperradial wave number, and $i$ and
$f$ label the initial and final channels, repectively. 
The present results were obtained using the mass of helium
atoms and the model dimer potential $v(r)=D\mbox{sech}^2(r/r_0)$
with $D$ and $r_0$ adjusted to give a single two-body $s$-wave
bound state.  

In the adiabatic hyperspherical representation, recombination for
$a>0$ is driven primarily by the broadly peaked nonadiabatic coupling  
between the lowest three-body entrance channel and the highest
molecular channel~\cite{Esry02}. In this picture, there are two indistinguishable  
pathways for recombination, leading to the so-called ``St\"uckelberg
oscillations''. This interference phenomena modifies the $a^4$
dependence of the rate, suppressing it for certain values of $a$.  At
zero energy, the analytic results predict these minima to be equally
spaced on a logarithmic scale and separated by a factor of
approximately $e^{\pi/\alpha}\approx 22.7$, where ${\alpha}=1.0064$.  

For $a<0$, the recombination rate is enhanced for particular values of
$a$ and, with the help of the adiabatic hyperspherical representation,
can be interpreted as three-body tunneling through a potential barrier
in the entrance channel~\cite{Esry02}. The nonadiabatic coupling is localized at
small $R$ behind this barrier so that recombination is suppressed for
energies below the barrier maximum. That is, unless the collision
energy matches the energy of a three-body resonance trapped behind
this barrier.  Under these conditions, transmission through the
barrier jumps and strong enhacement of the recombination rate can be
observed. As with the interference minima, these resonances
can be associated with Efi\-mov phy\-sics \cite{Efimov}, and are also
predict to be equally spaced on a logarithmic scale (separated by
a factor of about 22.7). 
\begin{figure*}[htbp]
\begin{center}
\includegraphics[width=2.25in,angle=270]{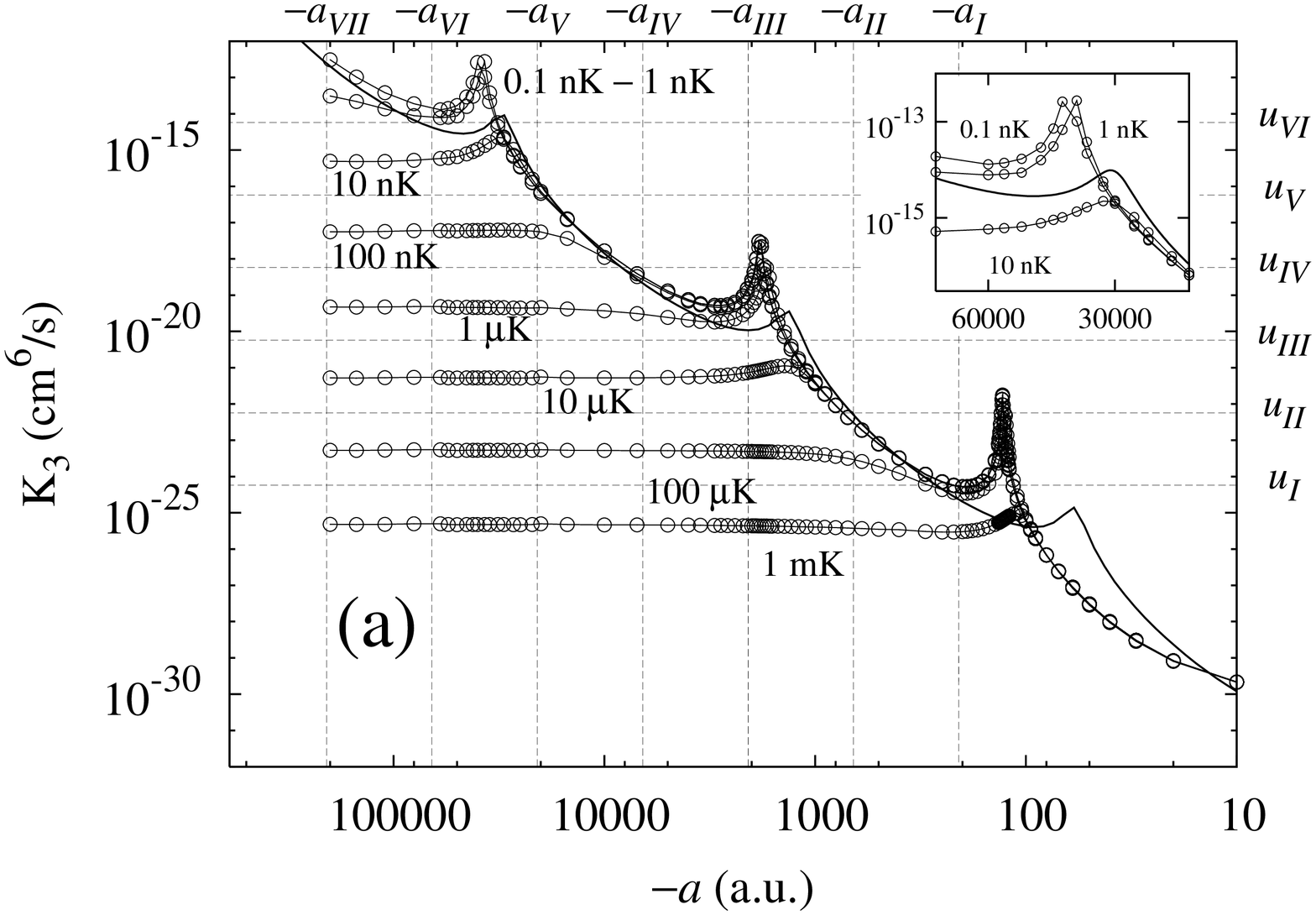} 
\includegraphics[width=2.25in,angle=270,clip=true]{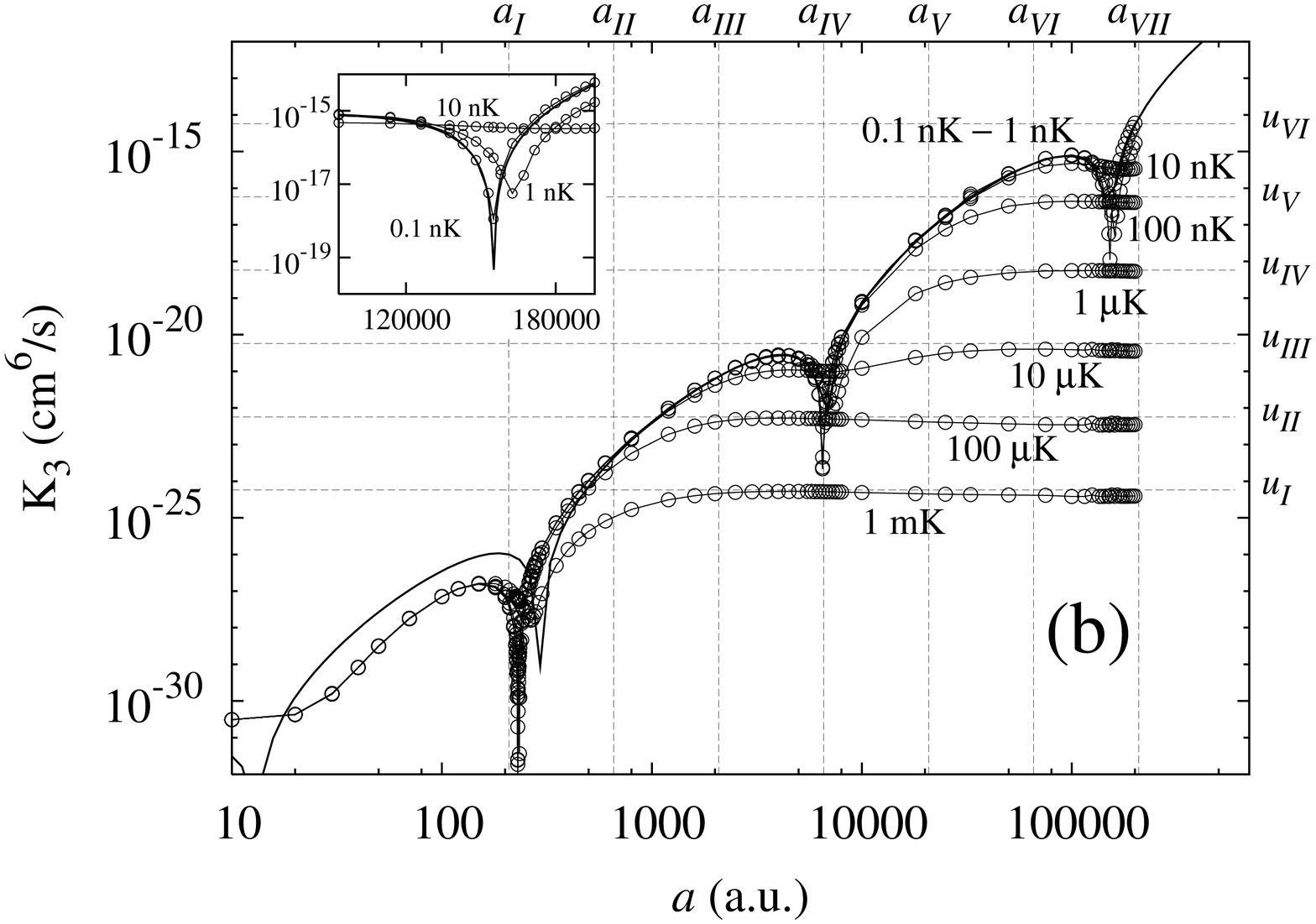}
\end{center}
\caption{Three-body recombination rate for (a) $a<0$ and (b) $a>0$.
The horizontal lines represent the unitarity limit $u_N$ for each energy
(reported as a temperature using $E=k_BT$).
The vertical lines represent the boundary of the threshold regime
$a_N$. The solid curve represents the analytical formula
Eq.~(\ref{K3Eqn}). The insets show the third resonance peak (a) and 
third interference minima (b) from $a=0$.}  
\label{recrateposneg} 
\end{figure*}
Figure~\ref{recrateposneg} shows the $J^{\pi}=0^{+}$ recombination
rates calculated at energies in the range 0.1~nK to 1~mK. 
Figure~\ref{recrateposneg}(a) shows the first three resonance peaks 
for $a<0$, and Fig.~\ref{recrateposneg}(b) shows the first three
interference minima for $a>0$. For small values of the scattering
length, the rates for all energies lie along a common, {\em universal}
curve. For any given energy, the rates depart from this universal curve
at some value of the scattering length, with the highest energies
departing soonest.

In the adiabatic hyperspherical representation, the analytic
recombination expression is derived under the assumption that
the collision energy is in the threshold regime.  It is natural,
then, to conclude that the analytic expression --- and thus
the universal behavior --- is only valid in the threshold regime.
The collision energy is in the threshold regime when it is smaller
than other characteristic energies of the system.  One obvious
energy scale is the two-body binding energy $E_{12}$.

In Fig. \ref{recrateposneg}, the vertical dashed lines 
mark the scattering lengths determined from the relation $E\lesssim
E_{12}$ for each energy. It is clear that the two-body binding energy 
provides a reasonable estimate for the domain of universal behavior,
i.e., for each energy, the rate curve for $a$ less than this limit
follows the common curve.  For $a<0$, a better, more restrictive limit
can be determined from the adiabatic potential since the threshold
regime in this case requires energies less than the potential barrier
maximum, $U_{\rm max}=0.079/\mu a^{2}$~\cite{Esry01}, which reduces
the limiting $a$ by about a factor of three. For example, in a
$^{23}$Na condensate at a temperature of 100~nK, the recombination
rate is expected to be universal only for --3200~a.u.$<a<8650$~a.u.;
for $^{87}$Rb, for --1650~a.u.$<a<4450$~a.u.  All of these values are
well within the range that are already experimentally accessed near
Feshbach resonances.

%

We also show in Fig.~\ref{recrateposneg} the analytical
results~\cite{Nielsen,Esry02,Braaten01,Braaten02}: 
\begin{equation}
K_3 = \begin{cases}
\frac{4590\:(a^4/m) \sinh(2\eta_*)}{\sin^2[\alpha \ln(3|a|/2r_{0})+\Phi+1.63]+\sinh^2\eta_*} & a<0, \\
         {\scriptstyle 360\:(a^4/m) \sin^{2}\left[\alpha\ln({3a}/{2r_{0}})+\Phi\right]} & a>0 
      \end{cases}
\label{K3Eqn}
\end{equation}
where $\Phi$ and $\eta_*$ are unknown parameters. $\Phi$
represents an unknown small-$R$ phase~\cite{Nielsen,Esry02} (related to 
$\Lambda_*$ in Refs.~\cite{Braaten01,Braaten02}) 
and is chosen to give the best fit to
the third interference minimum at 0.1~nK. The additional 1.63~rad of $a<0$ phase
is predicted in \cite{Braaten02}.  The value 
$\eta_{*}$=0.1 was found to give the best fit for $a<0$. There
is generally very good agreement with the numerical results for large,
positive $a/r_0$, and Eq.~(\ref{K3Eqn}) appears to be essentially exact for zero energy recombination. 
It relies on the effective range expansion, however, and 
gets increasingly worse as $|a|$ decreases due to order $r_{0}/a$
errors (here, $r_0$=15~a.u.).  The agreement is more qualitative for $a<0$
due to the small shift of the resonance peak positions.
We found, though, that a 15\% change in the extra $a<0$ phase gives
good agreement with the 0.1~nK curve.

%

One factor left out of Eq.~(\ref{K3Eqn}) is
unitarity (although a ``unitarized'' version has been proposed~\cite{Chris}
to help explain the experimental results in Ref.~\cite{Weber}). 
As the collision energy grows large compared
to $E_{12}$ for a fixed scattering length, the probability of
recombination approaches unity for the $0^+$ partial wave.  More
relevant for experiments, unit recombination
probability is also reached as the scattering length is increased at fixed
collision energy $E$. The horizontal dashed lines in
Fig.~\ref{recrateposneg} denote the unitarity limit ---
$u_{N}=192\pi^2/\mu k_{N}^{4}$, obtained from Eq.~(\ref{K3Def}) by
setting $|S|^2$=1 --- for each energy shown. From the figure, it is
clear that the recombination rate reaches the unitarity limit for
positive $a$ outside the threshold regime. For negative $a$, however,
while the rate does saturate, it does so at a value about a factor of
ten below unitarity. The main effect of unitarity is to restrict the
number of resonances or minima observable at a given energy.

%

\begin{figure}[htbp]
\begin{center}
\includegraphics[width=1.5in,height=1.6in,angle=270]{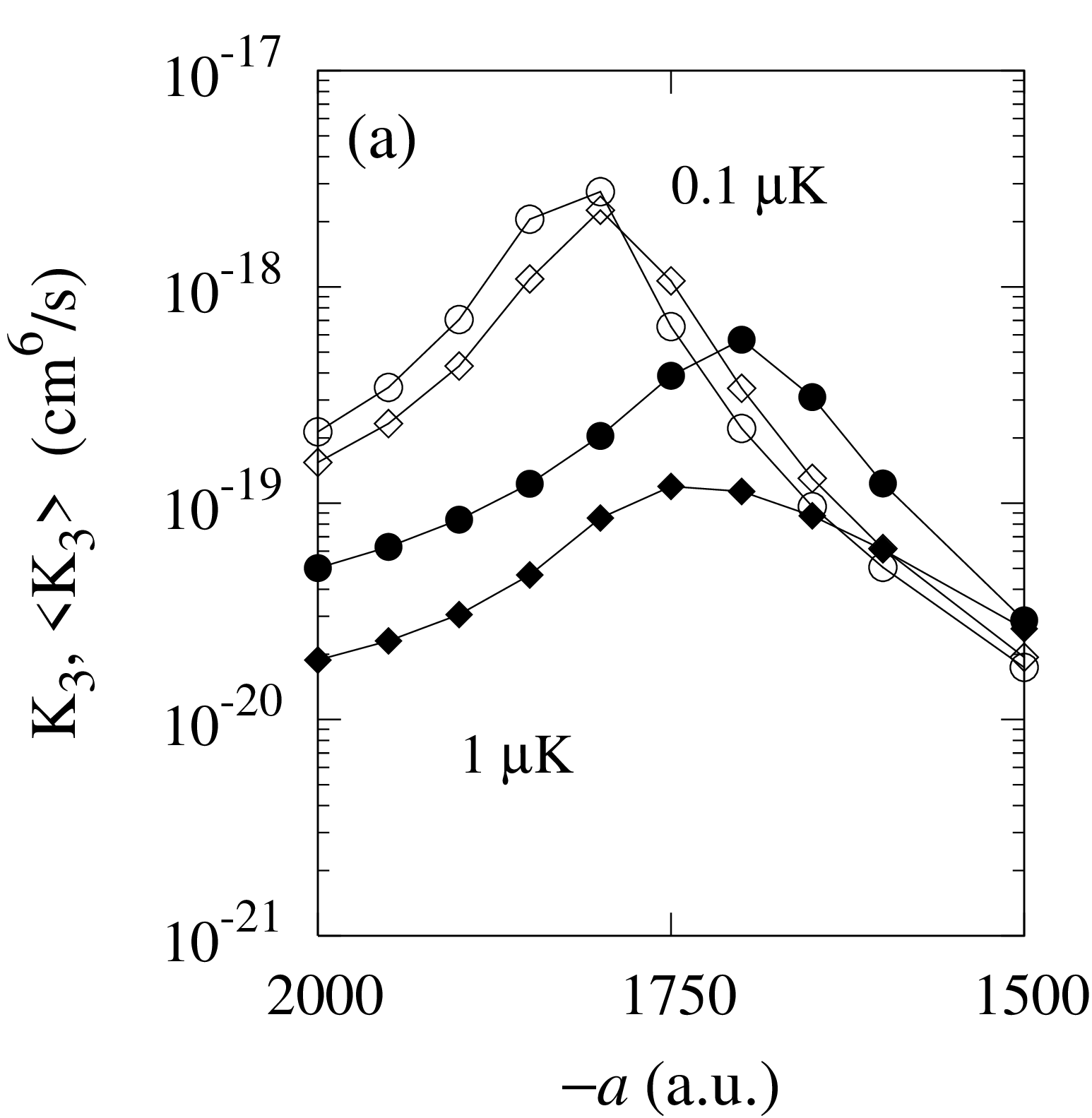}
\includegraphics[width=1.5in,height=1.6in,angle=270]{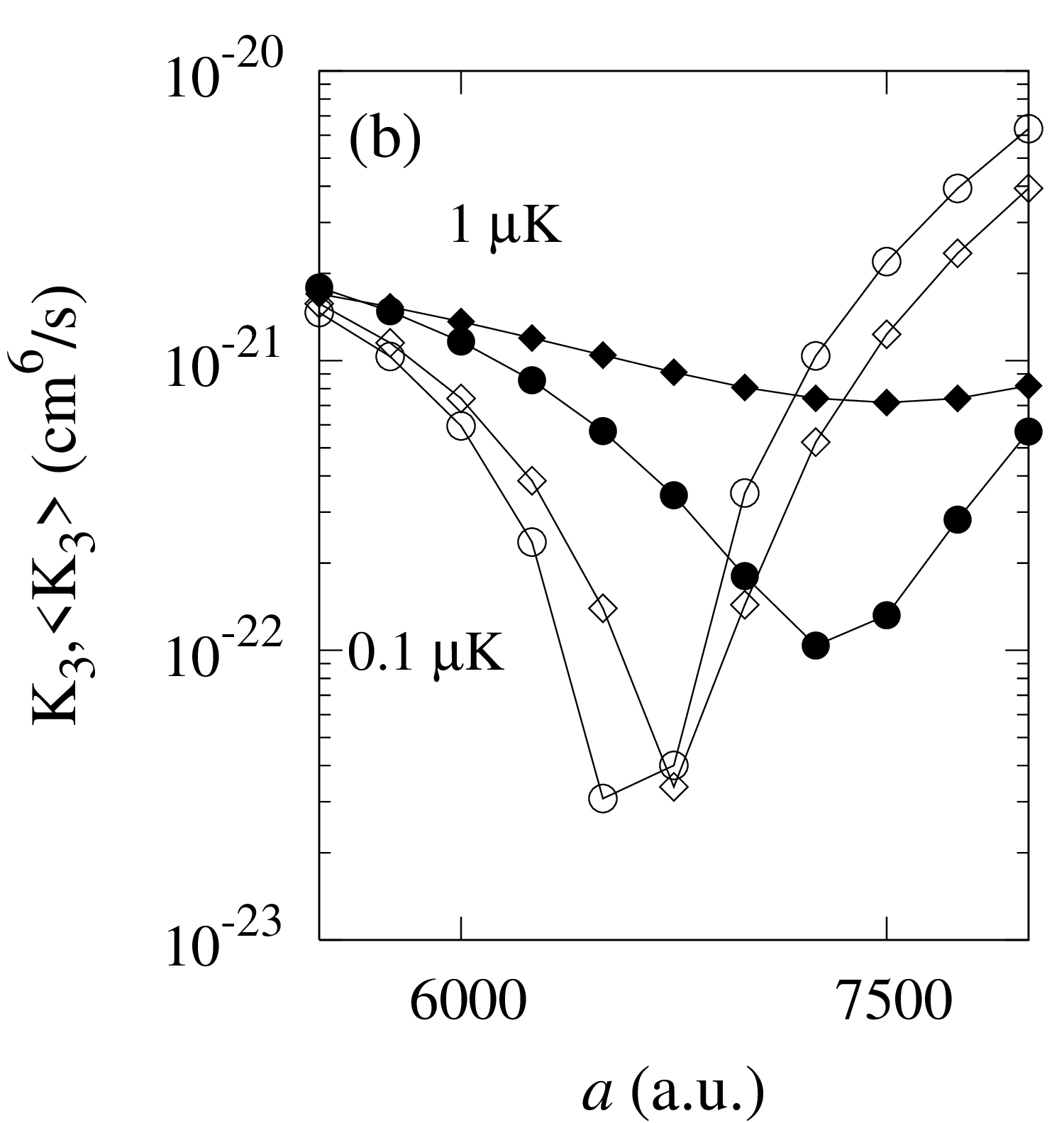}
\end{center}
\caption{
Thermally averaged recombination rate (a) near
the second resonance peak and (b) near the second interference
minimum. Circles and diamonds represent $K_3$ and $\langle
K_3\rangle$, respectively, at 0.1~$\mu$K (open symbols) and $1
{\mu}$K (filled symbols).}  
\label{thermalavgfig}     
\end{figure}

A second factor neglected in Eq.~(\ref{K3Eqn})
is the thermal average.  Experiments are performed at fixed
temperatures rather than fixed energies, so the thermal average
becomes crucial for proper comparison with experiment.  In the
threshold regime, the recombination rate is constant as a function of
energy, so the thermal average has no effect.  Since we consider
exactly the situation when the system is no longer in the threshold
regime, thermal averaging can have significant effects.  The thermally
averaged recombination rate is  
\begin{equation}
\langle K_{3} \rangle(T)=\frac{1}{2(k_{B}T)^{3}}\int K_{3}(E)E^{2}
e^{-E/k_{B}T}dE,
\label{thermalavg}
\end{equation}
\noindent
where $k_{B}$ is Boltzmann's constant.  Figure~\ref{thermalavgfig} 
illustrates the effects of thermal averaging at 0.1~${\mu}$K and
1~${\mu}$K near the second resonance peak and second interference
minima.  For energies solidly within the threshold regime, thermal
averaging has little effect. For energies on the border of the
threshold regime, however, averaging reduces the intensity of both the
peaks and minima, making their observation 
much more difficult.

%

\begin{figure}[htbp]
\begin{center}
\includegraphics[width=1.5in,height=1.6in,angle=270]{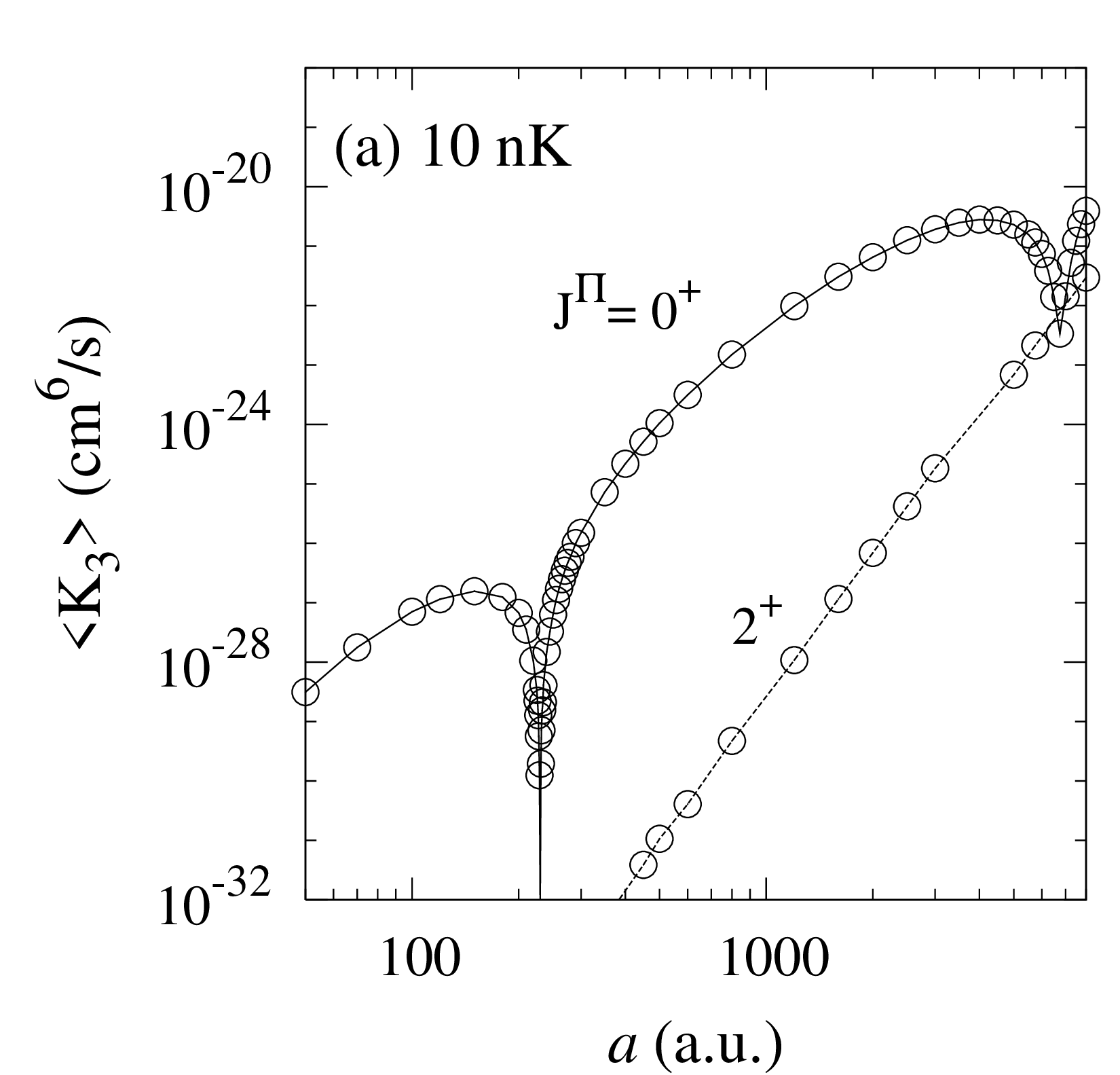}
\includegraphics[width=1.5in,height=1.6in,angle=270]{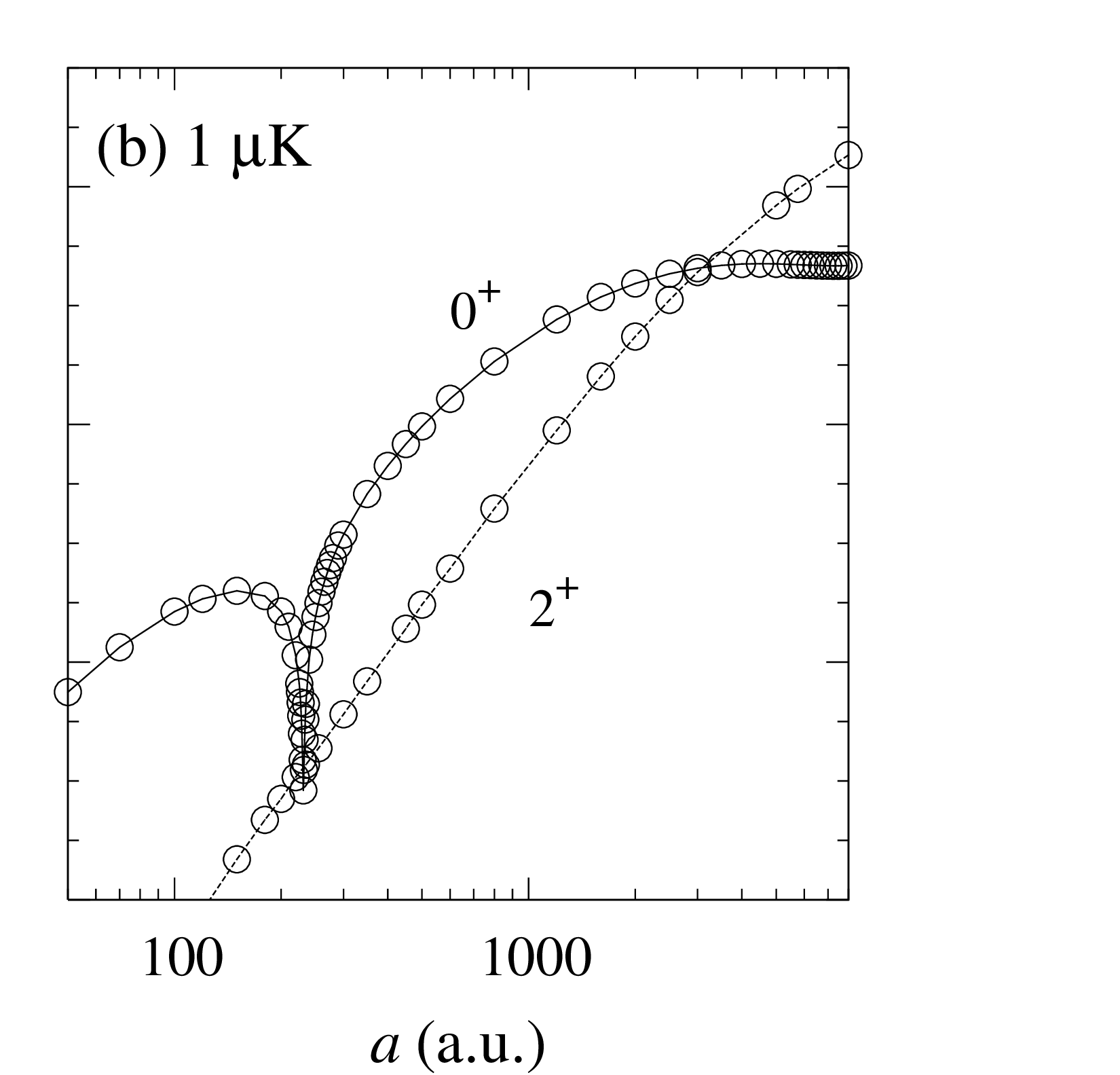}
\end{center}
\caption{Comparison of the $0^{+}$ and $2^{+}$ recombination
rate for $a>0$ and temperatures of (a) 10~nK and (b) 1~$\mu$K.}
\label{ratej0j2}     
\end{figure}

A third factor not included in Eq.~(\ref{K3Eqn}) is
the contribution from higher partial waves.  The $J^{\pi}=2^{+}$ rate
was calculated for --3000~a.u.$<a<$8000~a.u. and energies from 0.1~nK
up to 1~mK.  The 2$^+$ threshold law is $K_3\propto E^2
a^8$~\cite{Esry04}, so for a finite energy, there will be a scattering
length for which the 2$^+$ contribution is comparable to
$0^+$. Figure~\ref{ratej0j2} shows the thermally averaged $2^{+}$ rate at 10~nK and
1~$\mu$K along with $0^{+}$ for $a>0$.  (For $a<0$, the $2^{+}$
recombination rate is many orders of magnitude smaller than for $0^+$,
making it completely negligible for the present range of scattering
lengths and energies.) It is clear from the figure that the $2^{+}$
rate dominates $0^+$ at the second interference minimum for 1~$\mu$K
so that the total rate will show just one minimum. At 10~nK, the 2$^+$
rate is merely comparable to $0^+$ at the second minimum, cutting its
depth in the total rate. For energies below 10~nK, the $2^{+}$
recombination rate is negligible in this range of scattering length;
for larger scattering lengths, however, the $2^{+}$ recombination rate
can contribute substantially.   

Taken together, the unitarity limit, thermal averaging, and higher
partial waves restrict the analytic results for ultracold
three-body recombination to the threshold regime, i. e., when
$E\lesssim E_{12}$. 

Experimentally, the consequences are rather dramatic.  If we
imagine tuning the scattering length using a Feshbach resonance, 
then $a$ will, for instance, change from its background value to
$+\infty$, then to $-\infty$, then again to its background value ---
all while the system is at essentially the same temperature.  The
analytic expressions predict that the three-body recombination rate
grows like $a^4$ as the resonance is approached, goes through an
infinite number of minima as $a\rightarrow+\infty$, then has an
infinite number of resonances as $a$ returns from $-\infty$. Each
series of features reflects Efimov physics, so measuring them might
reveal evidence for this effect.  The present calculations
show, however, that the infinite series are truncated to a
small number ($\approx \alpha/\pi \ln(3a_{c}/2r_0)$, where
$a_c=\hbar/\sqrt{2\mu_{12}E_{12}}$) for typical 
experimental parameters and that the contrast of the surviving
features may be considerably reduced. The recombination rate is thus
not a good candidate for observing physics related to the Efimov
effect except at extremely low temperatures. 

Even though we have shown that the universal behavior described by 
existing analytic expressions is limited to the threshold regime, 
scattering lengths up to a few thousand atomic units are included.
Moreover, a new sort of universal behavior dictated by the unitarity
limit may take over and modifications to the analytic
expressions along these lines have already been
proposed~\cite{Chris}. Since we have used only one model potential, we
are not in a position to discuss any universal behavior outside
of the threshold regime.  We expect, however, that recombination for $a>0$ will
be much as we have shown in Fig.~\ref{recrateposneg} since it takes place
at large distances where differences in the two-body potential will have
little effect.  For $a<0$, the situation is just the opposite since
recombination is a small distance process.  The resonance positions
as well as the $a\rightarrow -\infty$ limiting rate will likely
then depend on the two-body potential.

\acknowledgments{This work was supported by the National
 Science Foundation and by the Research Corporation.}

\end{document}